\documentclass[useAMS,usenatbib]{mn2e}
\usepackage{epsfig}
\title[Search for gravitational lens candidates in the XMM-LSS/CFHTLS common field]
      {Search for gravitational lens candidates in the XMM-LSS/CFHTLS common field}

  \author[A. Elyiv et al.]
  { A.~Elyiv$^{1,2}$\thanks{E-mail: elyiv@astro.ulg.ac.be}, O. Melnyk$^{1,3}$, F. Finet$^{1}$, A. Pospieszalska-Surdej$^{1}$,
  L. Chiappetti$^{4}$,  \newauthor{M. Pierre$^{5}$,  T. Sadibekova$^{5}$ and J. Surdej$^{1}$}\\  
$^1$ Institut d'Astrophysique et de G\'eophysique,
     Universit\'e de Li\`ege, 4000 Li\`ege, Belgium  \\
$^2$Main Astronomical Observatory, Academy of Sciences of Ukraine, 27 Akademika Zabolotnoho St., 03680 Kyiv, Ukraine\\
$^3$Astronomical Observatory, Kyiv National University, 3 Observatorna St., 04053 Kyiv, Ukraine\\
$^4$ INAF, IASF Milano, via Bassini 15, I-20133 Milano, Italy\\
$^5$ Laboratoire AIM, CEA/ DSM/Irfu/SAp, CEA-Saclay, F-91191 Gif-sur-Yvette Cedex, France\\
}

\date{Released 2013 June 04}
\pagerange{\pageref{firstpage}--\pageref{lastpage}} \pubyear{2013}

\def\LaTeX{L\kern-.36em\raise.3ex\hbox{a}\kern-.15em
    T\kern-.1667em\lower.7ex\hbox{E}\kern-.125emX}

\begin{document}

\label{firstpage}

\maketitle

\begin{abstract}
{Our aim was to identify gravitational lens candidates among some 5500 optical counterparts of the X-ray point-like sources in the medium-deep $\sim 11$ sq. deg. XMM-LSS survey.
We have visually inspected the optical counterparts of each QSOs/AGN using CFHTLS T006 images. We have selected compact pairs and groups of sources which could be multiply imaged QSO/AGN. 
We have measured the colors and characterized the morphological types of the selected sources using the multiple PSF fitting technique. 

We found three good gravitational lens candidates: J021511.4-034306, J022234.3-031616 and J022607.0-040301  which consist of pairs 
of point-like sources having similar colors. On the basis of a color-color
diagram and X-ray properties we could verify that all these sources are good QSO/AGN candidates rather than stars. 
Additional secondary gravitational lens candidates are also reported.} 

\end{abstract}


\section{Introduction}

Gravitational lensing remains a powerful tool for studies of dark
matter distribution in the Universe. One important application of gravitational 
lens statistics is to provide strong constraints on the cosmological
models and parameters since the lensing optical depth depends on the cosmological volume element 
(Surdej et al. 1992, Mitchell et al. 2005, Oguri et al. 2008, Jullo et al. 2010, Rozo et al. 2011). 

An ideal set of observations to carry out extragalactic research and especially for searching gravitational lens systems is the CFHTLS photometric survey \footnote{Canada France Hawaii
Telescope Legacy Survey: http://www.cfht.hawaii.edu/Science/CFHLS/}. Previous investigators have mainly set up automatic procedures to search for gravitational lens systems. 
Cabanac et al. (2007) have presented 40 strong lens candidates over 28 sq. deg. of the CFHTLS field, which consist of galaxy/group deflectors producing conspicuous 
gravitational arc systems. More et al. (2012) have reported about 127 lens system candidates (Strong Lensing Legacy Survey-ARCS) over 150 sq. deg in the CFHTLS. 
In that work, they applied semi-automatic techniques to find gravitational arcs with the help of some additional visual inspection.
Maturi et al. (2013) studied the statistical color properties of gravitational arcs over 37 sq. deg. in the CFHTLS-Archive-Research Survey
and found 73 new arc candidates. Sygnet et al. (2010) have found 3 good candidates 
for edge-on galaxy lens applying their automatic/visual procedure to 30444 individual CFHTLS relatively bright spirals 18$<i<$21. 
Shan et al.(2012) have performed a weak lensing analysis over 64 sq. deg. of the CFHTLS fields.

We propose here to search for multiply imaged QSOs/AGN among the counterparts of X-ray point-like sources detected with XMM.
Therefore contrary to the previous approaches, we have been searching for gravitationally lensed QSOs/AGN at much smaller angular scales (typically $1''$-$4''$).

The search for gravitational lens candidates among the optical counterparts of X-ray selected QSOs/AGN looked
a priori very promising as, first of all, it allows the selection of a sample of
QSOs/AGN with very few contaminants. Moreover, as the X-ray QSOs/AGN are bright and distant sources (observed up to a redshift $z\sim 4$),
they have a high probability of being multiply imaged due to the presence of a deflector near their line-of-sight.

From lensing probability calculations over a mock catalogue of X-ray sources (reproducing the redshift distribution and
differential number counts of the COSMOS X-ray sources as a function of their flux), where we have assumed a constant comoving
density of deflector galaxies modelled as singular isothermal spheres, we expect $\sim 15$ QSOs/AGN among the XMM-LSS soft X-ray
sources to have a deflector sufficiently close to their line-of-sight to lead to the formation of multiple lensed images of
their optical counterpart (Surdej et al. 2011, Finet et al. 2012).

The contiguous $\sim 11$ deg$^2$ medium-deep XMM-LSS field has been surveyed at high galactic latitudes with the XMM-Newton satellite, 
centered on RA $2^{h}18^{m}00^{s}$ DEC $-7^{\circ} 00'00''$ (J2000). It provides the identification of a homogeneous sample of galaxy 
clusters and QSOs/AGN (Pierre et al. 2011) and reaches a sensitivity
near $10^{-18}$~W~m$^{-2}$ for point-like sources in the soft band (Elyiv et al. 2012).
About $80\%$ of the 6000 point-like XMM-LSS X-ray sources have (an) optical CFHTLS DR6 counterpart(s) (Chiappetti et al. 2013). 
Visual inspection of all counterparts led us to classify the sources as being either extended or point-like and reject the 
wrong associations (Melnyk et al. 2013). In this paper we describe separately the results of our gravitational lens
search among the CFHTLS counterparts of X-ray sources.

In Section 2 we detail the method used to identify multiply imaged QSOs/AGN in the XMM-LSS field.
In Section 3 we describe the adopted procedure to search for multiply imaged QSOs/AGN, 
we discuss each individual candidate and report their color measurements as well as their morphology.
Section 4 contains the main conclusions.



\section{The sample of gravitational lens candidates}

Our sample of X-ray point-like sources and their optical counterparts consist of objects which have
been detected in the soft and/or the hard X-ray band and coming from the multiwavelength catalogue of Chiappetti et al. (2013). 
The $\sim 5500$ optical counterparts of the X-ray sources were visually inspected on $10''\times 10''$ CFHTLS 
optical images centered on the X-ray sources.
The angular separations between the lensed images for a typical galaxy deflector and cosmological redshifts of the source (QSO/AGN) 
and the lens are typically of the order of 1 or several arcsec 
 (Refsdal and Surdej 1994). Knowing that lensed components should be point-like sources and should have rather similar color properties (color difference smaller 
than $0.1-0.2$~mag.), 
this selection was also carried out using these strict criteria.
Contamination by the lens, intrinsic color variation plus time
delay(s) and/or micro-lensing acting may cause color differences above typical CFHTLS photometric errors.

The PSF of the XMM-Newton telescopes provides a resolution near $4''-6''$ for a
point-like source. A detected point-like X-ray source may therefore be sometimes identified
with multiple (more than one) optical counterparts. We have made an in-depth study of these
particular cases in order to detect new gravitational lens systems (mostly multiply imaged QSOs/AGN).

\begin{figure*}
\begin{tabular}{c c c c c}

\epsfig{file=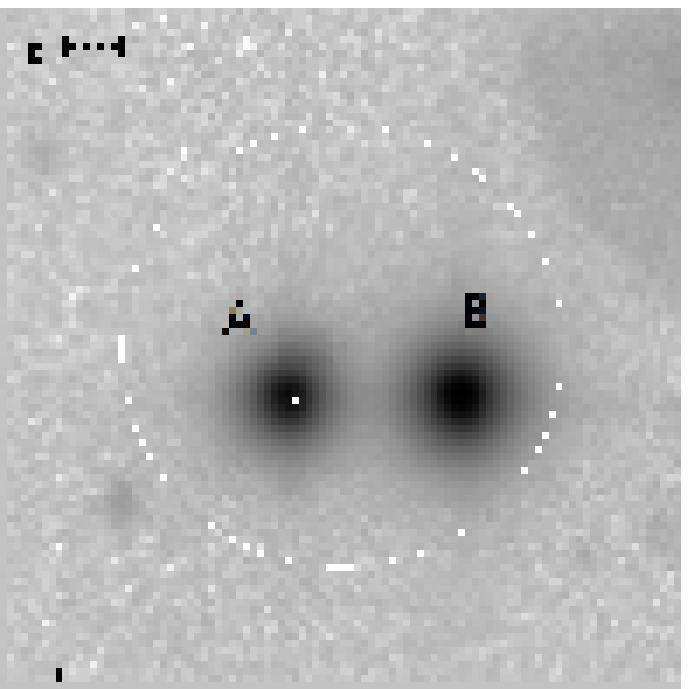,width=3.3cm}&\epsfig{file=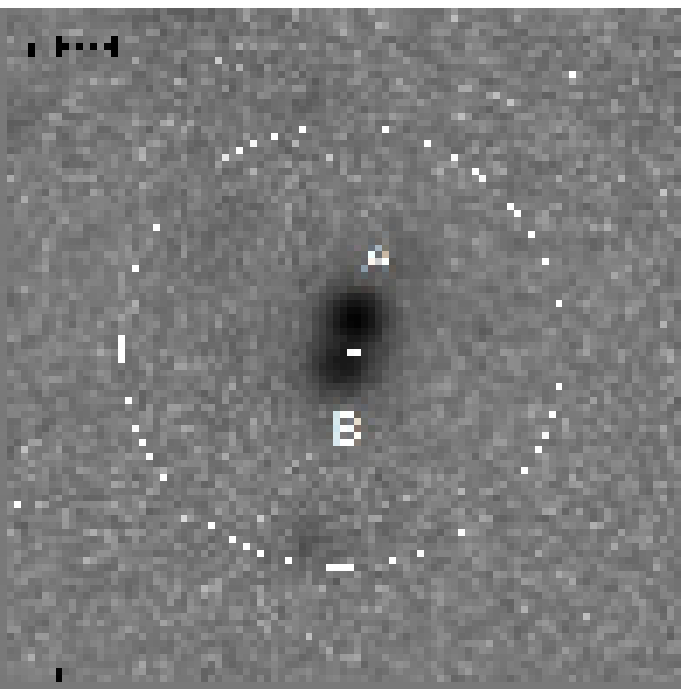,width=3.3cm}&\epsfig{file=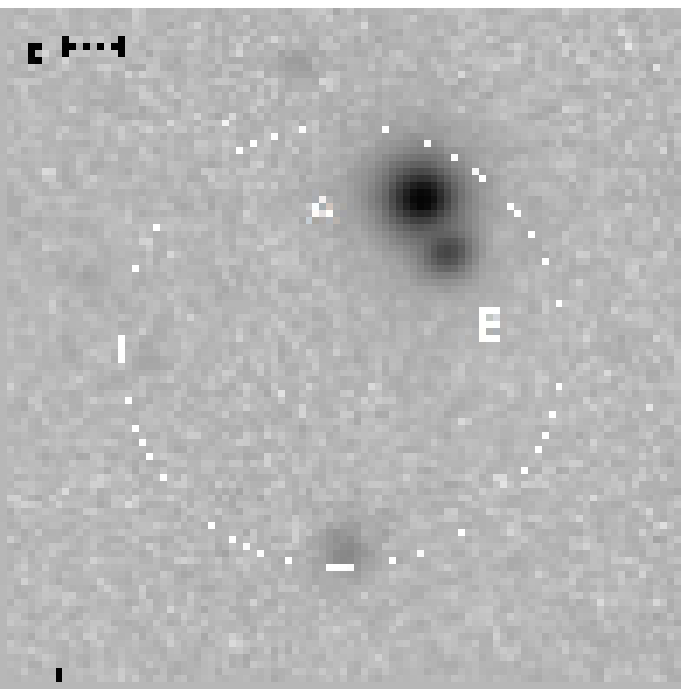,width=3.3cm}&\epsfig{file=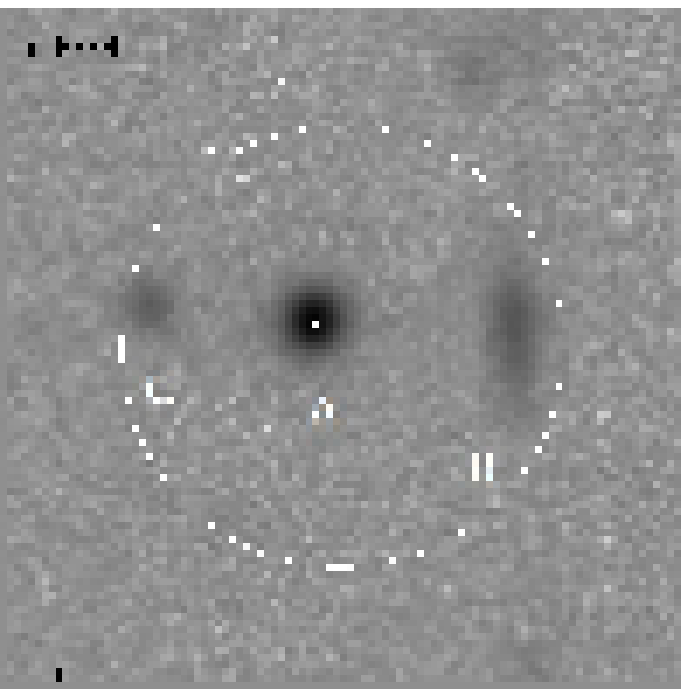,width=3.3cm}&\epsfig{file=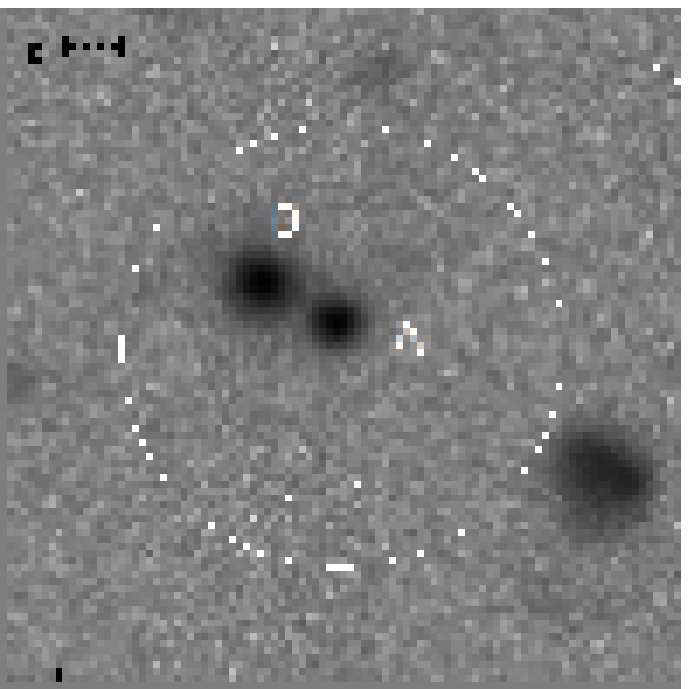,width=3.3cm}\\
J021511.4-034306               &J022234.3-031616               &J022607.0-040301               &J021844.4-044825 &J021936.7-055721\\

\epsfig{file=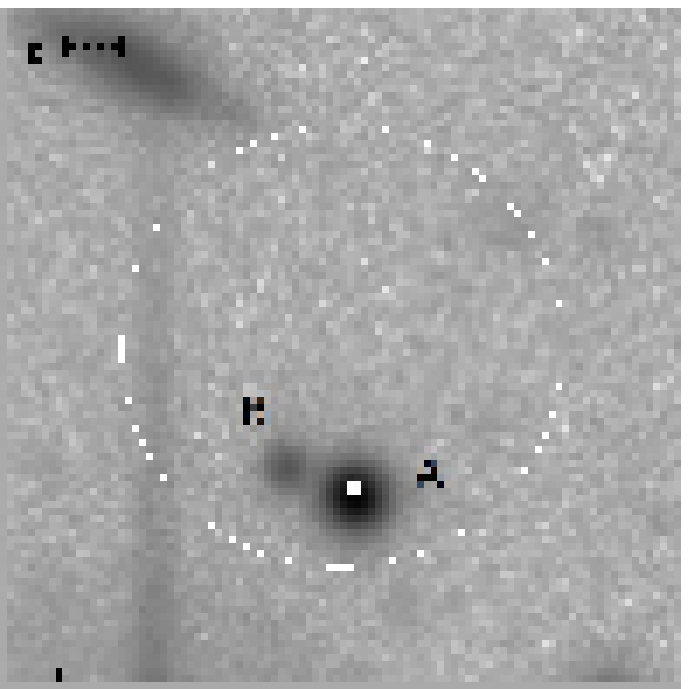,width=3.3cm}&\epsfig{file=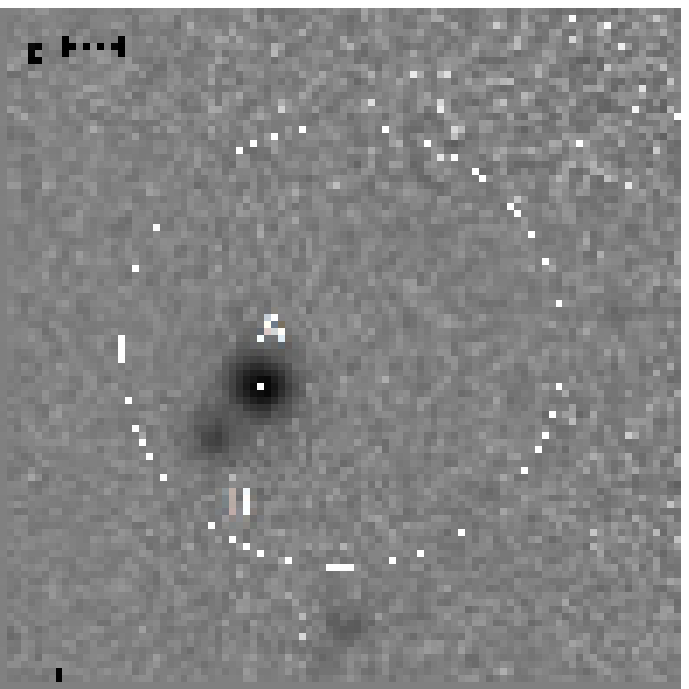,width=3.3cm}&\epsfig{file=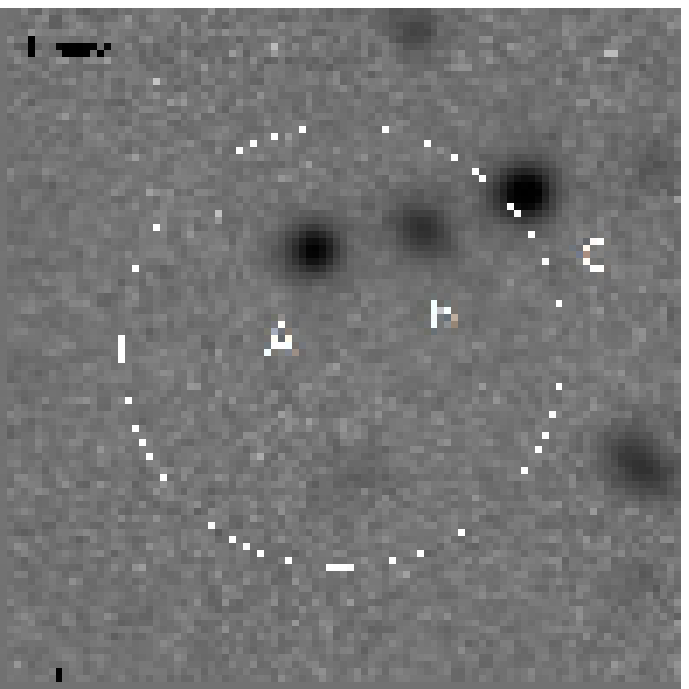,width=3.3cm}&\epsfig{file=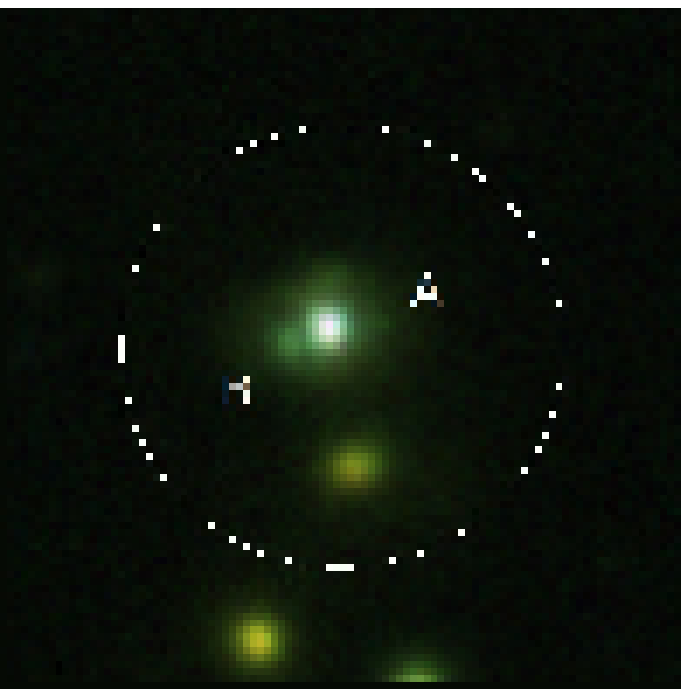,width=3.3cm}&\epsfig{file=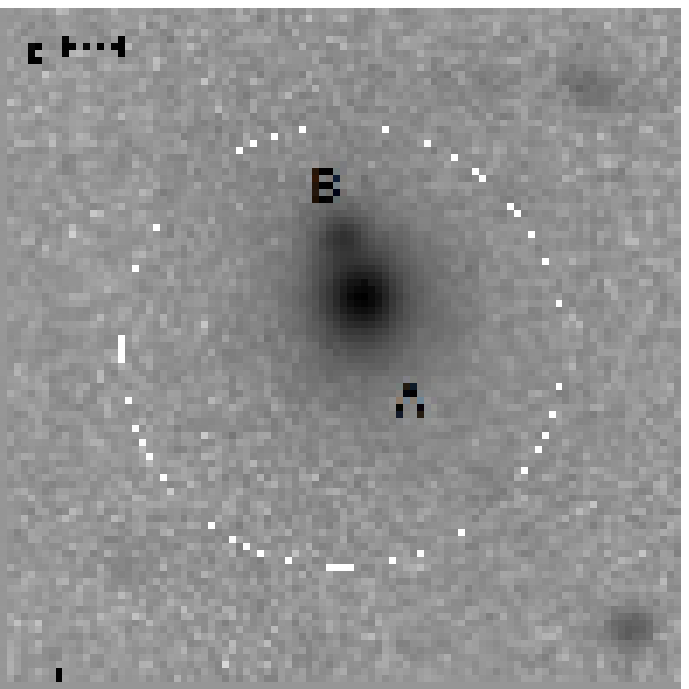,width=3.3cm}\\
J022055.1-060132               &J022500.2-052204               &J022324.0-054928               &J022739.7-050044               &J022509.7-050950 \\

\epsfig{file=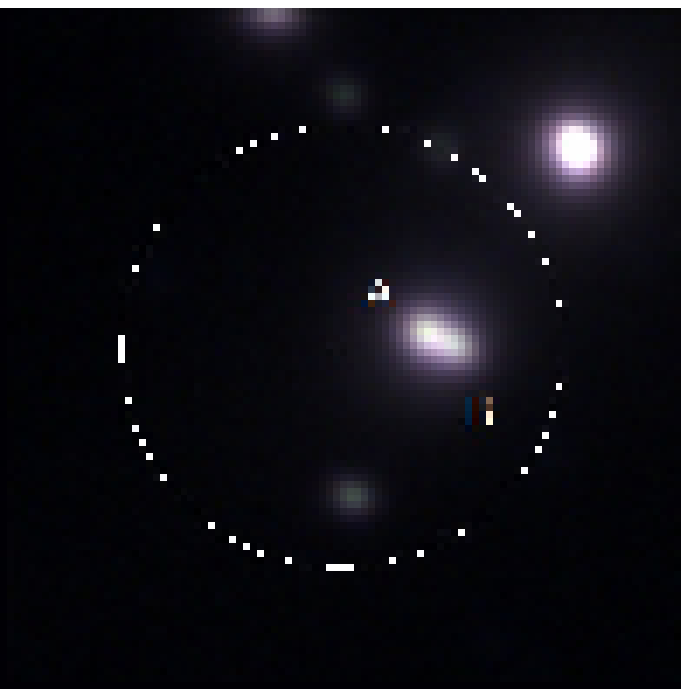,width=3.3cm}&\epsfig{file=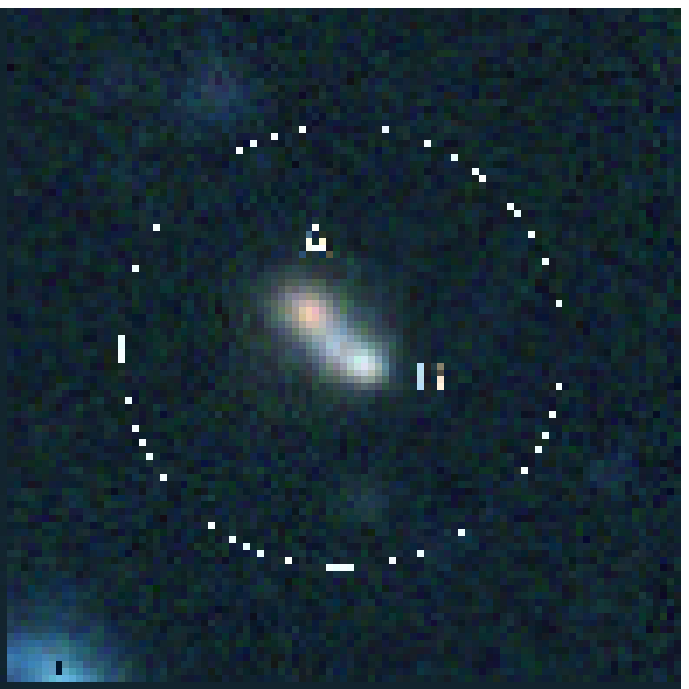,width=3.3cm}&\epsfig{file=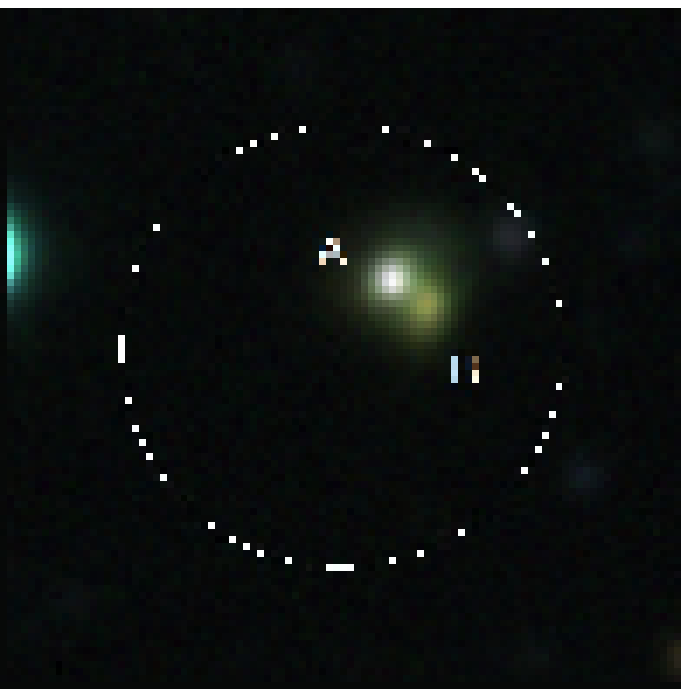,width=3.3cm}&\epsfig{file=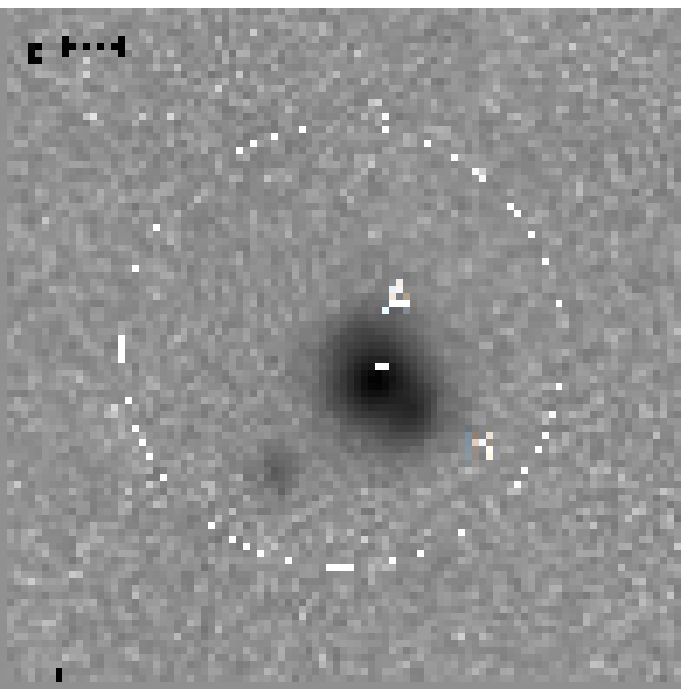,width=3.3cm}&\epsfig{file=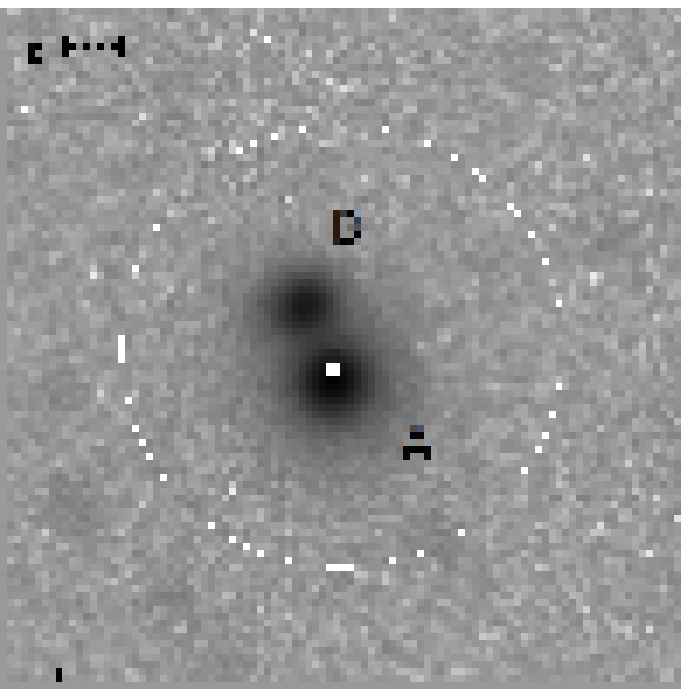,width=3.3cm}\\
J021956.2-052809               &J021929.6-040856               &J021736.4-041944               &J021947.4-041034               &J021649.2-054327\\

\epsfig{file=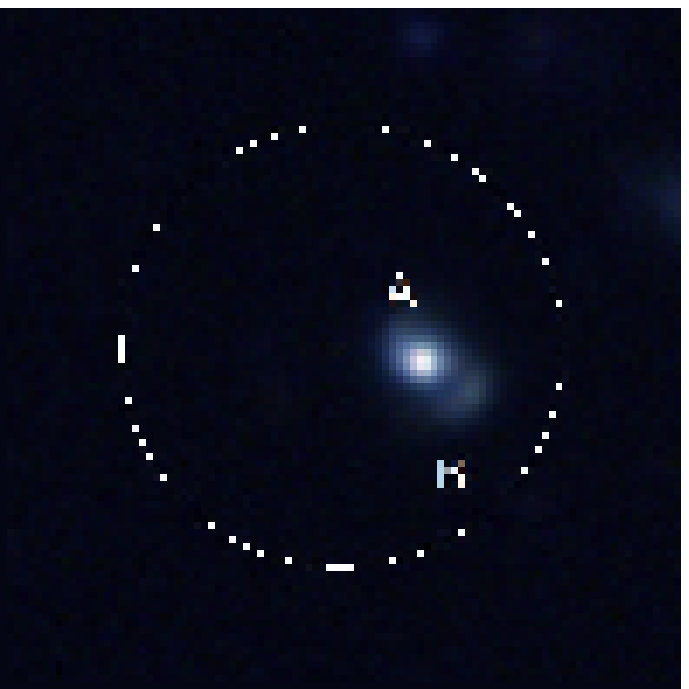,width=3.3cm}&\epsfig{file=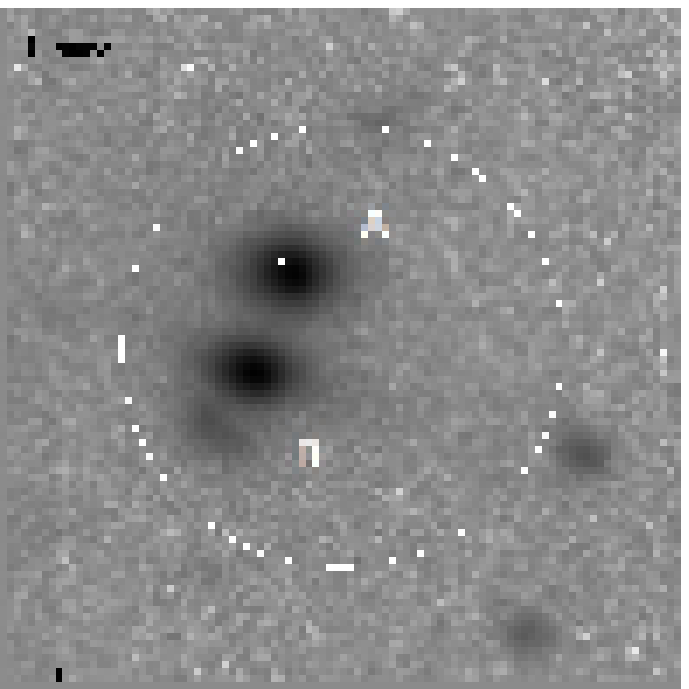,width=3.3cm}&\epsfig{file=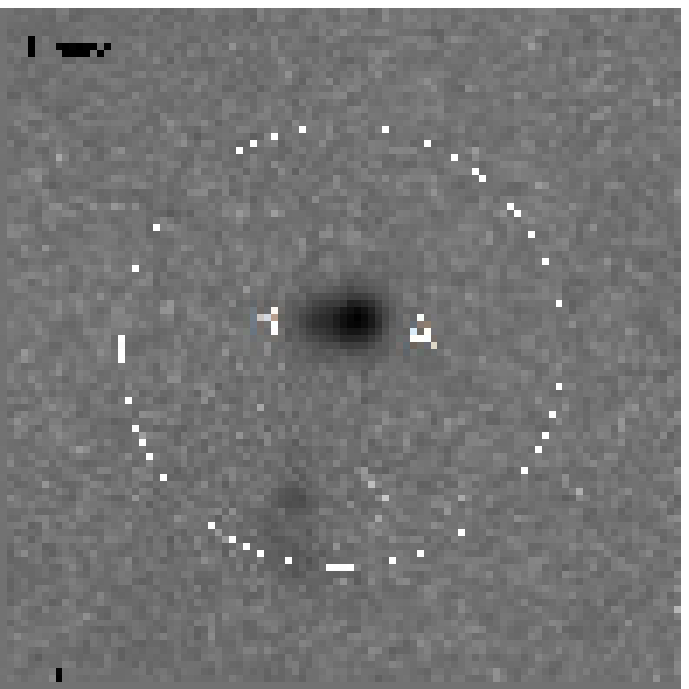,width=3.3cm}\\
J021902.8-054109               &J021634.8-043302               &J022646.2-045156\\

\end{tabular}
\caption{Images of 18 gravitational lens candidates. The cross indicates
the center of the X-ray emission. The radius of the circle around the center
of X-ray emission is $6''$.
Here we chose the best illustrative band or color image for each
candidate.}
\label{fig1}
\end{figure*}

We inspected {\it g},~{\it r} and {\it i} CFHTLS direct CCD frames whenever they were available. For visual
inspection we used the monochromatic images in logarithmic scale 
for each band as well as the color composed images. For better objectivity four independent
members of the present team (called hereafter ``inspectors``) ranked each counterpart in the following way:
$1$ - very good candidate, $2$ - good lens candidate, $3$ - possible
lens candidate and $0$ - definitely not a candidate. 
A similar approach has been followed by More et al. (2008) when searching for gravitational lenses in the Extended Chandra Deep Field South and 
by Jackson (2008) in the COSMOS field.  
We finally retained nearly 300 candidates which were marked with the 1, 2 
or 3 rank by at least one inspector.
Next we added the ranks from all inspectors and formed a top list of
72 probable lens candidates. We visually inspected this list several times and removed very faint
objects, objects which are obvious stars according to their spectra and so on. 
At the end, we finally selected 18 multiply imaged QSO/AGN candidates which could be potentially strong lens systems, see Fig.~\ref{fig1} and Table~1.
They are ranked there from the best to worst according to final morphological and color selection.

Most of them consist of double sources. For 5 pairs with an angular
separation larger than $3''$ we considered their photometry 
in the $g$,~$r$ and $i$ bands directly from the CFHTLS catalog. We used
available spectroscopic or photometric redshifts from SDSS \footnote{www.sdss.org} and Melnyk et
al. (2013). For the remaining more compact systems we performed a multiple PSF
fitting analysis to precise their morphology (point-like or extended
source) and accurate magnitude measurement of each component. We adopted the
same PSF fitting technique as that used for the analysis of the
gravitational lens systems HE 0435-1223 described in Ricci et al. (2011) and UM673 in
Ricci et al. (2013). 

\begin{table*}
\caption{Properties of the potential strong lens candidates}
\label{table:1}
\begin{tabular}{c c c c c c c c c c}
\hline\hline
ID & & Xcatname$^{(1)}$ & R.A. & DEC. & $g$ & $r$ & $i$ & $g-r$ & $r-i$ \\
\hline
1&A$^{(5)}$&J021511.4-034306&$2:15:11.49$&$-3:43:07.94$&$18.15$&$17.76$&$17.76\pm0.001$&0.39&0.00\\
&B$^{(5)}$&&$-4.69''$&$+0.03''$&$17.24$&$16.86$&$16.78$&0.38&0.08\\
\hline
2&A$^{(5)}$&J022234.3-031616&$2:22:34.33$&$-3:16:17.14$&$21.59\pm 0.02$&$22.05\pm 0.03$&&$-0.46$&\\
&B$^{(5)}$&& $+0.37''$&$-1.25''$ &$22.28\pm 0.03$&$22.64\pm 0.04$&&$-0.36$&\\
\hline
3&A$^{(5)}$&J022607.0-040301&$2:26:06.86$&$-4:02:57.09$&$19.35\pm 0.01$&$19.08 \pm 0.01$&$19.00\pm 0.01$&0.27&0.08\\
&B$^{(5)}$&&$-1.39''$& $-2.95''$& $21.82\pm 0.02$&$21.47\pm 0.03$&$21.18\pm 0.02$&0.35&0.29\\
\hline
4&A$^{(5)}$&J021844.4-044825&$2:18:44.46$&$-4:48:24.69$&$22.62\pm0.01$&$20.81\pm0.01$&$19.98\pm0.002$&1.81&0.83\\
&B$^{(2)}$&&$-5.41''$&$-0.16''$&$22.97\pm0.03$&$22.19\pm0.03$&$21.66\pm0.02$&0.78&0.53\\
&C$^{(2)}$&&$+4.51''$&$+0.44''$&$24.11\pm0.05$&$23.17\pm0.05$&$22.52\pm0.03$&0.94&0.65\\
\hline
5&A$^{(5)}$&J021936.7-055721&$2:19:36.80$&$-5:57:20.26$&$22.83\pm0.02$&$22.24\pm0.02$&$22.09\pm0.02$&0.59&0.15\\
&B$^{(2)}$&&$+2.00''$&$+1.03''$&$22.21\pm0.01$&$21.55\pm0.01$&$21.37\pm0.01$&0.66&0.18\\
\hline
6&A$^{(5)}$&J022055.1-060132&$2:20:55.17$&$-6:01:36.51$&$21.10\pm 0.01$&$20.44\pm 0.01$&$20.67\pm 0.02$&0.66&-0.23\\
&B$^{(5)}$&&$+1.86''$&$+0.77''$&$23.55\pm 0.06$&$23.47\pm 0.07$&$23.43\pm 0.06$&0.08&0.04\\
\hline
7&A$^{(5)}$&J022500.2-052204&$2:25:00.41$&$-5:22:05.50$&$22.20\pm 0.02$&$21.73\pm 0.02$&$21.35\pm 0.03$&0.47&0.38\\
&B$^{(5)}$&&$+1.17''$&$-1.38''$ &$23.90\pm 0.05$&$23.55\pm 0.06$&$23.59\pm 0.07$&0.35&-0.04\\
\hline
8&A$^{(2)}$&J022324.0-054928&$2:23:24.12$&$-5:49:25.13$&$21.69\pm0.004$&$21.64\pm0.01$&$21.32\pm0.01$&0.05&0.32\\
&B$^{(5)}$&&$-3.05''$&$+0.56''$&$23.11\pm0.02$&$22.74\pm0.04$&$22.06\pm0.02$&0.37&0.68\\
&C$^{(5)}$&&$-5.76''$&$+1.53''$&$21.05\pm0.003$&$20.83\pm0.01$&$20.76\pm0.01$&0.22&0.07\\
\hline
9&$^{(2)}$&J022739.7-050044&$2:27:39.83$&$-5:00:42.99$&&&\\
\hline
10&$^{(2)}$&J022509.7-050950&$2:25:09.72$&$-5:09:49.21$&&&\\
\hline
11&$^{(2)}$&J021956.2-052809&$2:19:56.12$&$-5:28:08.28$&&&\\
\hline
12&$^{(2)}$&J021929.6-040856&$2:19:29.66$&$-4:08:56.75$&&&\\
\hline
13&$^{(2)}$&J021736.4-041944&$2:17:36.35$&$-4:19:43.20$&&&\\
\hline
14&$^{(2)}$&J021947.4-041034&$2:19:47.40$&$-4:10:34.83$&&&\\
\hline
15&$^{(2)}$&J021649.2-054327&$2:16:49.22$&$-5:43:28.45$&&&\\
\hline
16&$^{(2)}$&J021902.8-054109&$2:19:02.66$&$-5:41:10.13$&&&\\
\hline
17&A$^{(2)}$&J021634.8-043302&$2:16:34.93$&$-4:33:00.13$&$21.93\pm0.01$&$20.52\pm0.01$&$19.93\pm0.004$&1.41&0.59\\
&B$^{(2)}$&&$+1.14''$&$-2.82''$&$21.77\pm0.01$&$20.44\pm0.01$&$19.82\pm0.01$&1.33&0.62\\
\hline
18&&J022646.2-045156&$2:26:46.26$&$-4:51:55.45$&$21.37\pm 0.02^{(3)}$&$^{(4)}$&$^{(2)}$&\\
\hline

\end{tabular}
\\ Remarks: \\
$^{(1)}$ Identifiers are from Chiappetti et al. (2013). In order to shorten the ID names, we systematically omitted the prefix 2XLSSd before the names of the objects. 
$^{(2)}$ Extended source. $^{(3)}$ Only for the bright A component. $^{(4)}$ Possibly affected by a cosmic ray. $^{(5)}$ Point-like source\\

\end{table*}

\section{Description of the individual candidates}

\subsection{The best candidates}

Candidate {\bf J021511.4-034306} is seen as a wide pair. Therefore we
took the magnitudes of the two point-like sources from the CFHTLS
database. The components show similar colors: $g-r=0.39$ and $0.38$
for the A and B components, $r-i=0.00$ and $0.08$, respectively. 
Photometric redshift determination for the A component indicates z=$0.94$ (Melnyk et al. 2013). We did
not find any SDSS spectra but just a note that these two sources
are point-like. This system thus consists of a good gravitational lens
candidate according to the PSF analysis. The components are bright and slightly
saturated. Therefore we encountered some problems when performing the
PSF fitting.


Candidate {\bf J022234.3-031616} consists of two close faint
sources. We have just {\it r} and {\it g}-band observations.
We performed a PSF fitting analysis and found that this system consists
of two point-like sources with very similar colors:
$g-r=-0.46$ and $-0.36$ for A and B, respectively, see Fig.~2. This system thus consists of a good gravitational
lens candidate.

Candidate {\bf J022607.0-040301} consists of one bright and one faint
source. We performed a PSF fitting analysis and found that both of them are
point-like sources.
However they have slightly different colors. Note that the presence of a nearby lens, micro-lensing and/or intrinsic variations with time delay acting
could of course account for such different colors. Photo-z for the A component is $0.086$ (Rowan-Robinson et al. 2008).

\subsection{Candidates found with extended images or with different colors}

Candidate {\bf J021844.4-044825} contains one bright optical component
near the center of the X-ray emission and two symmetric faint
components. The system is wide and we took the magnitudes of each component
from the CFHTLS catalog. The colors of the two symmetric sources are similar: $g-r=0.78$ and 0.94 for the B and C
components, $r-i=0.53$ and 0.65, respectively. Akiyama et al. (2013 in prep.) on the basis of spectroscopic observations 
claim that component A is a QSO/AGN with $z=4.55$. 
Since the central component A has very high redshift we do not consider it as a potential deflector.
Moreover the symmetric components B and C are extended, the system J021844.4-044825 does not therefore constitute good gravitational lens candidate.

Candidate {\bf J021936.7-055721} consists of two components with an
angular separation near $3''$. 
For these, we took the magnitudes from the CFHTLS database. The components
have similar colors: $g-r=0.59$ and $0.66$
for the A and B components, $r-i=0.15$ and $0.18$, respectively. We
performed a PSF fitting analysis of this system.
It shows that component A is point-like and that B is an extended
source.

Candidate {\bf J022055.1-060132} also consists of one bright and one faint
components. We performed PSF fitting and found that both components 
are point-like sources.
However they also show very different colors. 
Given the very different colors and high flux ratio, we do not consider
this system to be a good gravitational lens candidate.

Candidate {\bf J022500.2-052204} consists of one bright and one faint
components.
We performed PSF fitting and found that both components 
are point-like sources.
However they have very different colors. 
Given the very different colors and high flux ratio, we do not consider
this system to be a good gravitational lens candidate.

Candidate {\bf J022324.0-054928} is composed of one faint and two symmetric
bright components. The system is wide and we took the magnitudes of the components from the
CFHTLS catalogue. The colors of those components are very different.




\begin{figure}
\begin{tabular}{c c}
\epsfig{file=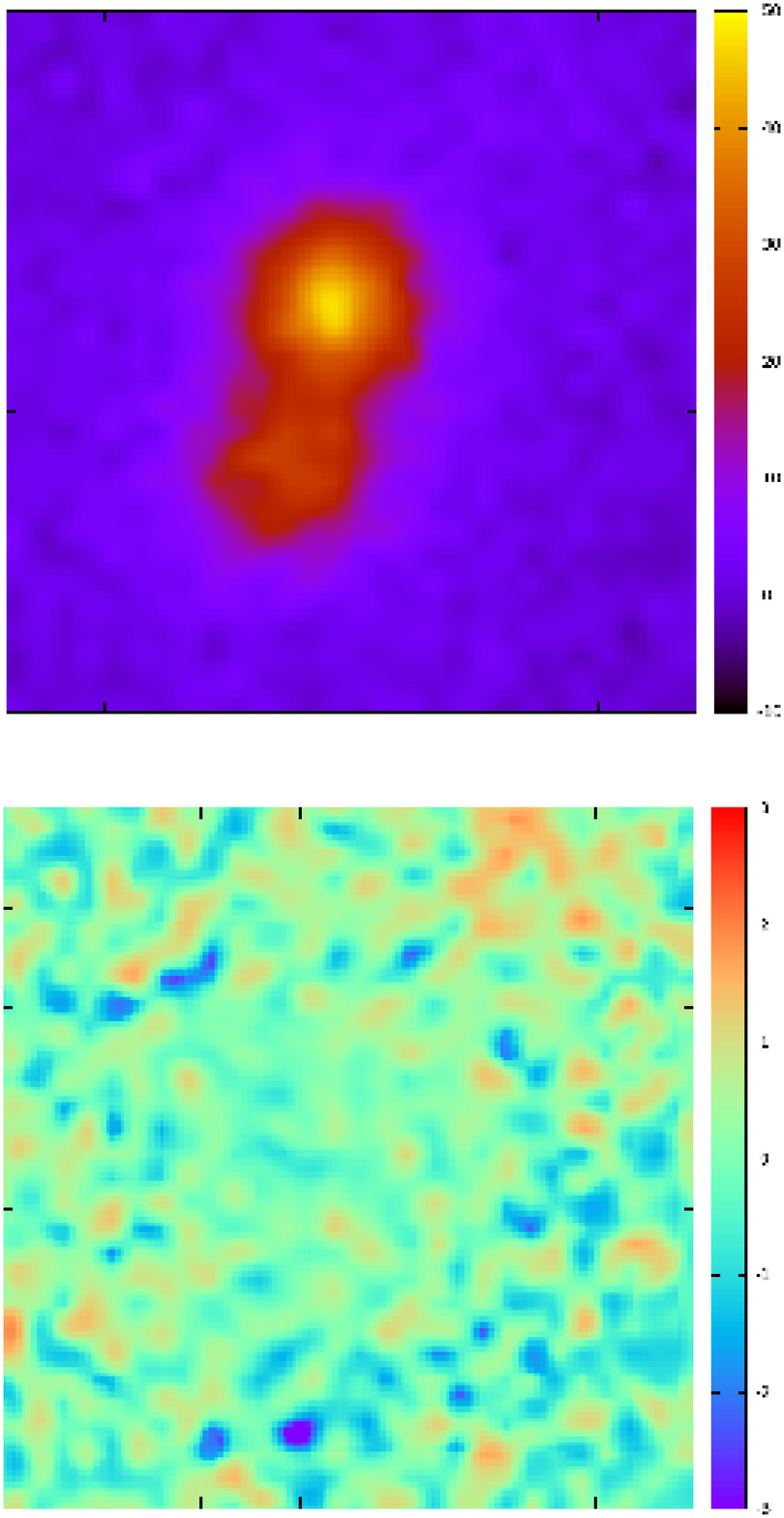,width=4cm}&\epsfig{file=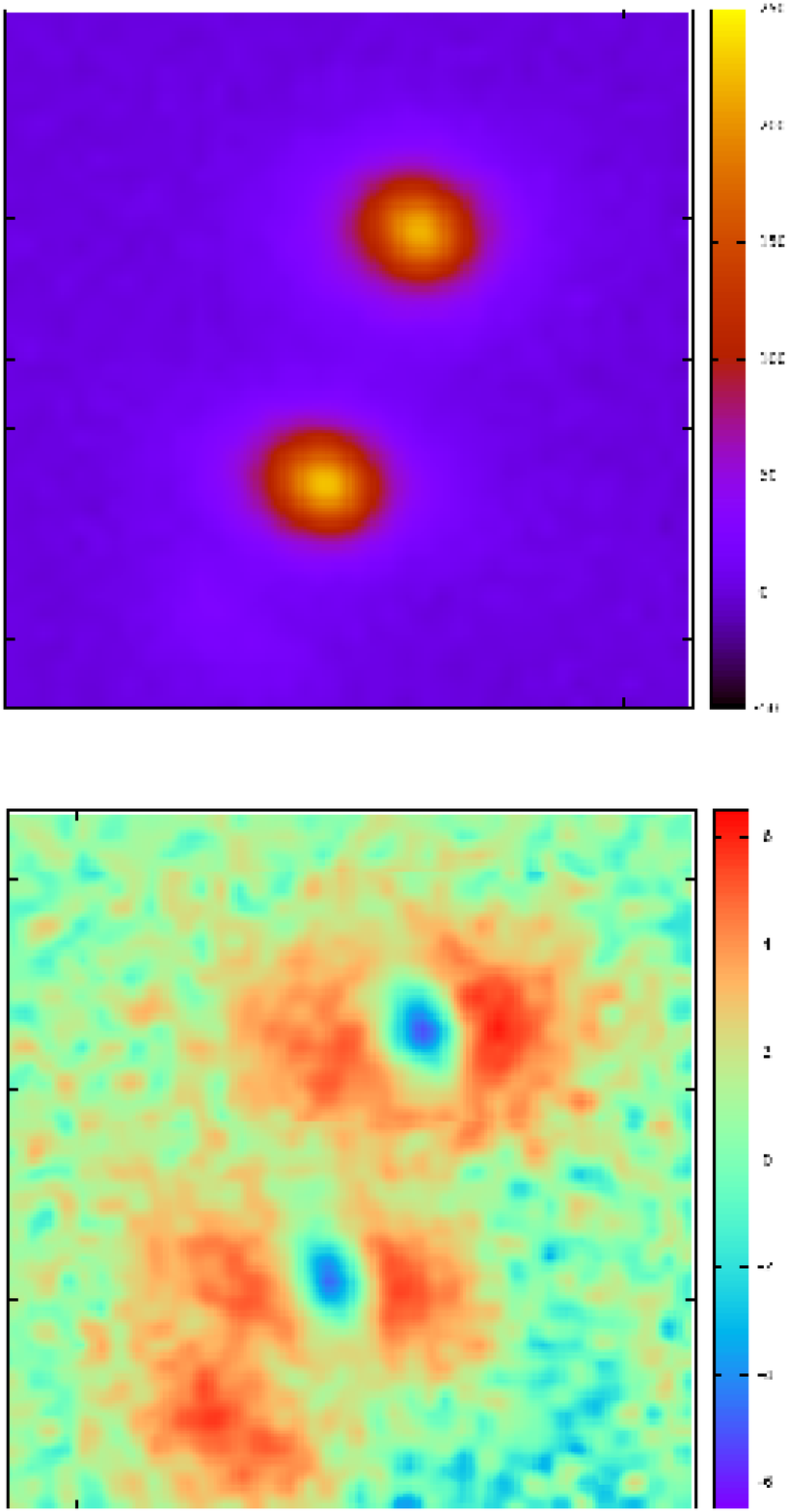,width=4cm}\\
{\bf J022234.3-031616} in the i-band & {\bf J021634.8-043302} in the i-band 
\end{tabular}
\caption{The CFHTLS images are shown at the top and the residuals after subtraction
of a double PSF at the bottom. The latter residuals have been normalized to the noise image, in $\sigma$ unit.}
\label{figJ021634.8-043302}
\end{figure}



\subsection{Candidates found to be extended or not classifiable sources}

Candidates {\bf J022739.7-050044, J022509.7-050950, J021956.2-052809,
J021929.6-040856, J021736.4-041944, J021947.4-041034, J021649.2-054327,
J021902.8-054109, J021634.8-043302 and J022646.2-045156} consist of pairs of extended sources according to our
PSF fitting analysis.

Candidate {\bf J021634.8-043302} consists of a wide pair and we took
their magnitudes from the CFHTLS data base.
The components show similar colors: $g-r=1.41$ and $1.33$
for the A and B components, $r-i=0.59$ and $0.62$, respectively.
From the SDSS database, the estimated photo-z is $0.38\pm0.04$ and $0.33\pm0.04$
for the A and B components, respectively. In the SDSS database there is just a note that the two components are
galaxies. We performed the PSF fitting of these two components and confirm that
they are both extended, see Fig.~\ref{figJ021634.8-043302}.

Candidate {\bf J022646.2-045156} looks very different in the various
bands. 
In the g-band it looks like a single point-like source according to the PSF fitting results.
In the r-band we see a faint secondary component. However the bright
component looks more compact than the reference star (PSF).
This could hint on the presence of a defect or cosmic ray in the r-band.
It was therefore not possible to correctly define the {\it r} magnitude.
In the i-band the bright component appears as an extended source
according to PSF fitting.
The SDSS SED analysis shows that it is a QSO/AGN with z=1.085. 
It could be that the host galaxy of this QSO/AGN has been detected in the i-band.


\begin{figure}
\epsfig{file=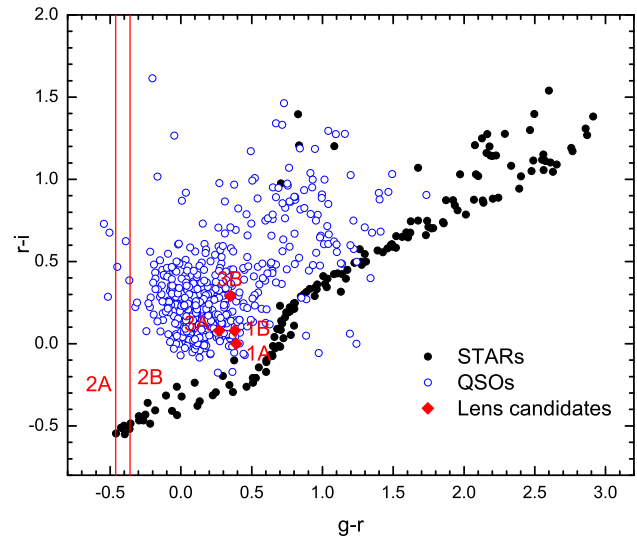,width=10cm}
\caption{Color-color diagram from the templates of stars and
spectroscopically confirmed QSOs/AGN in the XMM-LSS field,
131 templates are from Pickles A. (1998),
4 from Bohlin et al. (1995),
19 from Bixler et al. (1991). The components of the three best lens candidates are indicated in red: 
1A - J021511.4-034306A, 1B - J021511.4-034306B, 2A - J022234.3-031616A, 2B - J022234.3-031616B, 3A - J022607.0-040301A, 3B - J022607.0-040301B}
\label{fig_color}
\end{figure}

\subsection{Discussion}

Using the multiple PSF fitting technique, we found 11 gravitational lens system candidates which consist of extended
components: J021844.4-044825, J022739.7-050044, J022509.7-050950, J021956.2-052809,
J021929.6-040856, J021736.4-041944, J021947.4-041034, J021649.2-054327,
J021902.8-054109, J021634.8-043302 and J022646.2-045156.
The candidate J021936.7-055721 consists of two components which have similar
colors but one of the components is extended and the other one is point-like.
Using magnitudes from the CFHTLS survey or from PSF fitting for the case of
compact systems, we found that the following candidates have very different colors: J022055.1-060132, J022500.2-052204 and J022324.0-054928. 
 
The two candidates J021511.4-034306 and J022234.3-031616 consist of pairs of
point-like sources having very similar colors.
Candidate J021511.4-034306 is very bright in X-ray, $1.1\times 10^{-13}$ and $1.5\times 10^{-16}$~W~m$^{-2}$ in the soft and hard bands, respectively.
It corresponds to a hardness ratio $HR=-0.60$ which points to a source being likely
an unobscured AGN.
Candidate J022234.3-031616 is faint in X-ray and is only seen in the soft band, $6.8\times 10^{-18}$~W~m$^{-2}$.  
The optical counterparts of the J022607.0-040301 candidate have slightly
different colors but it could be caused by the presence of a lens, intrinsic color variation of the AGN combined with a time delay and/or microlensing effects. The system is bright in X-ray, about $1.5\times 10^{-17}$~W~m$^{-2}$ 
in both the soft and hard bands. Its corresponding hardness ratio $-0.63$ indicates
that it is a likely unobscured AGN.  
 
These last three systems (J021511.4-034306, J022234.3-031616 and J022607.0-040301) constitute the best gravitational lens candidates.

From the XMM-LSS source catalog (Chiappetti et al. 2013) we have constructed the color-color diagram ($r-i$ vs. $g-r$; see Fig. 3) for all spectroscopically confirmed stars 
and QSO/AGN counterparts of XMM-LSS point-like sources (see classification of sources in Melnyk et al. 2013). 
From this plot we suggest that the J021511.4-034306 and J022607.0-040301 candidates are probably AGN. 
For the candidate J022234.3-031616 we have only magnitudes in $g$ and $r$ bands
 and we cannot classify them as QSOs/AGN or stars on the sole basis of the color-color diagram.

\section{Conclusion}
We have performed a visual inspection of some 5500 optical counterparts of X-ray point-like sources in the XMM-LSS/CFHTLS common field.
Four independent inspectors have identified 18 compact systems which could be multiply imaged QSO/AGN. 
Using the PSF fitting technique we have characterized the morphology and colors of the components. 
As a result, we found {\bf three} systems (J021511.4-034306, J022234.3-031616 and J022607.0-040301) 
which met our color and morphological criteria and 
are the best gravitational lens candidates. They are point-like source pairs with similar colors. 
With a high probability they could be QSO/AGN. 
For a final confirmation we need spectroscopic observations of the multiple components of the selected systems.

\section*{Acknowledgments}
AE (post-doc PRODEX) and JS acknowledge support the ESA Prodex Programme ``XMM-LSS``, from the Belgian Federal Science Policy Office and
from the Communaut\'e{} fran\c caise de Belgique - Actions de recherche concert\'ees - Acad\'emie universitaire 
Wallonie-Europe. AE is grateful to Stephen Gwyn for help with the CFHTLS images.

\label{lastpage}

\end{document}